\documentclass{PoS}

\title{Impact of Theory Uncertainties on the Precision of the Top Quark Mass in a Threshold Scan at Future $e^+e^-$ Colliders}

\ShortTitle{Impact of Theory Uncertainties on Top Mass Measurements at Future $e^+e^-$ Colliders}

\author{\speaker{Frank Simon}\\
        Max-Planck-Institute for Physics, Munich, Germany\\
        E-mail: \email{fsimon@mpp.mpg.de}}

\abstract{Future energy-frontier electron-positron colliders will be capable of high-precision studies of top quark properties. The measurement of the top-pair production cross section around the threshold provides access to the mass of the top quark in theoretically well-defined schemes, with statistical uncertainties of 20 MeV or less, depending on the assumed integrated luminosity of the measurement. At this level of precision, experimental and theory systematics are likely to become important or even dominant. This contribution presents a first analysis of the impact of the remaining uncertainties of the recently completed calculation of the top pair production cross section at NNNLO QCD including the exchange of Higgs bosons on the extraction of the top quark mass from a threshold scan. The analysis is based on reconstruction efficiencies and background levels obtained in full simulation studies for CLIC, combined with signal cross sections from the higher-order calculations. To assess possible differences between different collider options, the study is performed in the context of CLIC, ILC and FCC-ee by considering the differences in the luminosity spectra.}

\FullConference{38th International Conference on High Energy Physics\\
		3-10 August 2016\\
		Chicago, USA}

\begin{document}

\section{Future $e^+e^-$ colliders and the $t\bar{t}$ threshold}

The precise measurement of top quark properties is one of the pillars of the physics program at future high-energy electron positron colliders. One of those properties is the top quark mass, which, together with measurements of the mass of the Higgs boson and the $W$ boson as well as coupling constants allows to make stringent tests of the Standard Model and enables the exploration of the stability of the Standard Model vacuum. 

The highest precision for the top quark mass is expected from a scan of the pair-production threshold in $e^+e^-$ collisions, which, in contrast to "conventional" measurements at the LHC, provide the mass in a theoretically well-defined framework, eliminating the interpretation uncertainties associated with the use of MC generator masses which are present otherwise. 

\begin{figure}[!ht]
\includegraphics[width=0.49\textwidth]{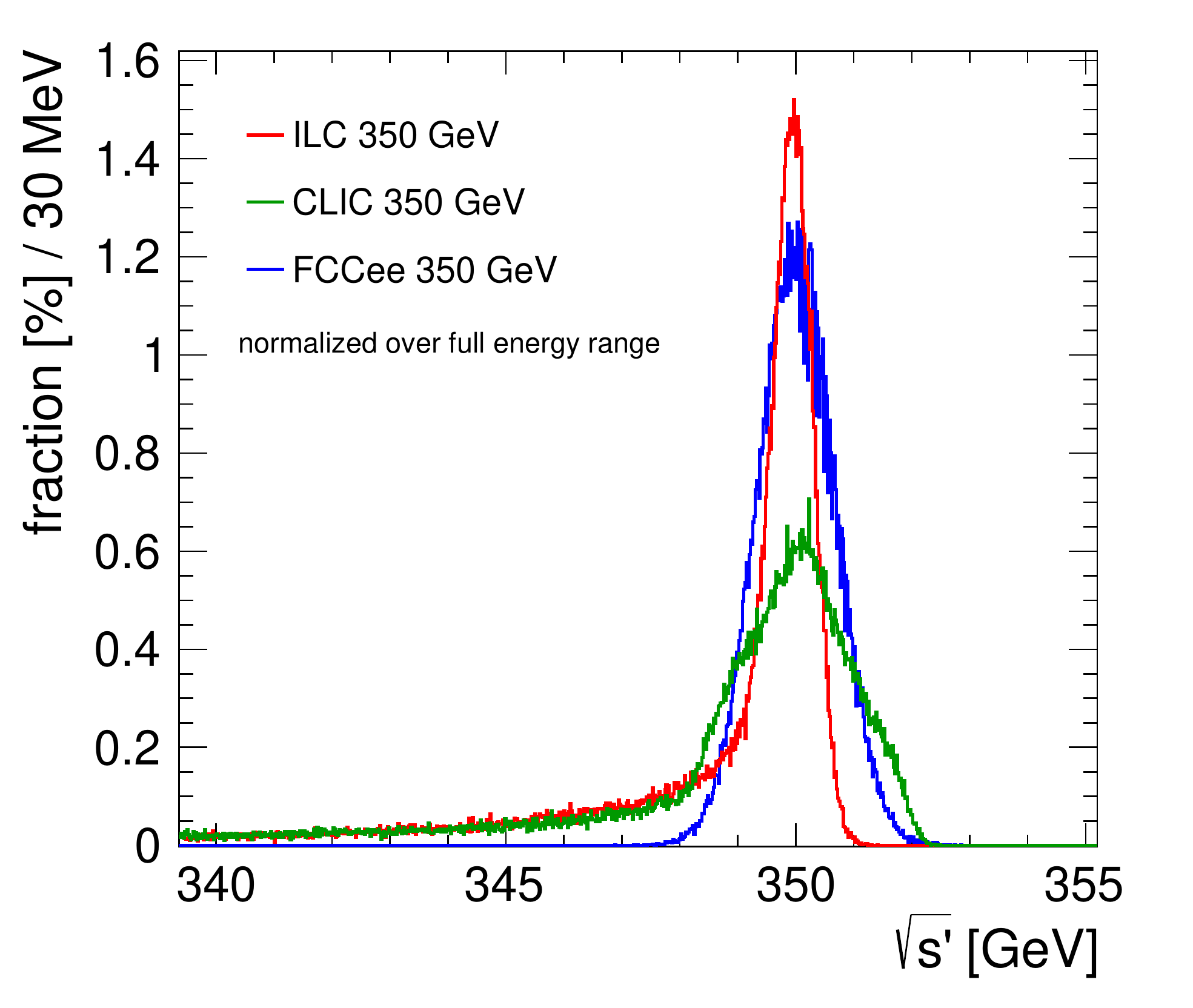}
\hfill
\includegraphics[width=0.49\textwidth]{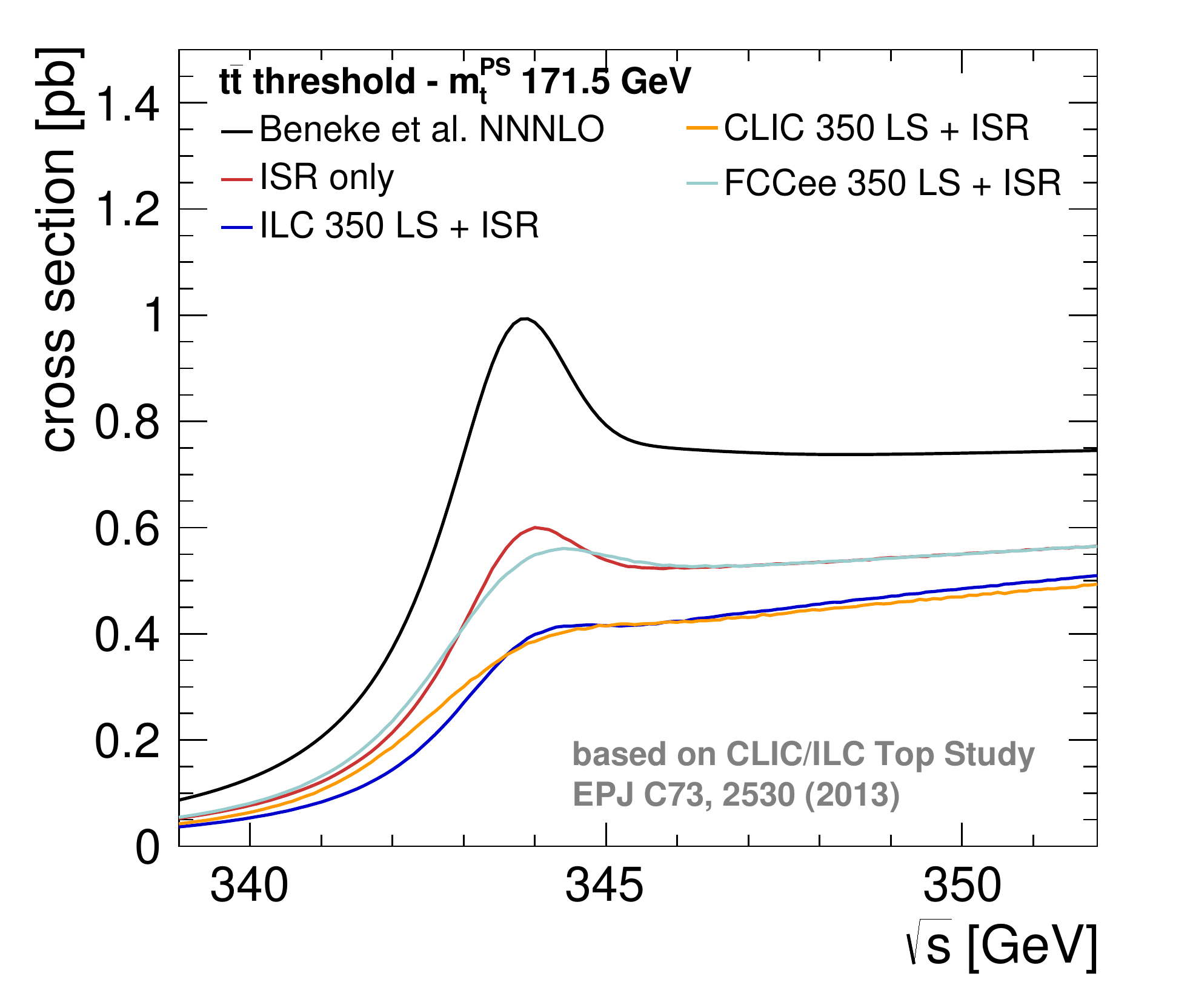}
\caption{The luminosity spectra of future high-energy $e^+e^-$ colliders ({\it left}) and the impact of these on the top pair production cross section in the threshold region for unpolarized beams ({\it right}). The FCC-ee luminosity spectrum is assumed to be a gaussian with a $\sigma$ of 0.19\%, based on the energy uncertainty quoted in \cite{Gomez-Ceballos:2013zzn}, while the ILC and CLIC luminosity spectra are taken from machine simulations. \label{fig:Lumi}}
\end{figure}

At present, three $e^+e^-$ colliders capable of reaching the energy required for top quark pair production are discussed as possible future large-scale projects in high energy physics: The International Linear Collider ILC \cite{Behnke:2013xla}, currently under consideration for hosting in Japan, the Compact Linear Collider CLIC \cite{Lebrun:2012hj} and the $e^+e^-$ Future Circular Collider FCC-ee \cite{Gomez-Ceballos:2013zzn} as part of the FCC project, both as possible future CERN facilities. All three of these machines provide a luminosity of 1 to 1.5 $\times$ 10$^{34}$cm$^{-2}$s$^{-1}$ at 350 GeV (with the possibility of experiments at up to 4 interaction points at FCC-ee), thus the main difference with regards to the precision of a top threshold scan is expected to be the luminosity spectrum, when assuming identical detector performance and ignoring the capabilities of the linear colliders for high beam polarizations. At each machine, a threshold scan with ten energy points with an integrated luminosity of 10 fb$^{-1}$ each could be performed within approximately one year at design luminosity.

Figure \ref{fig:Lumi} shows the luminosity spectra of the three different colliders at a center-of-mass energy of 350 GeV as well as the consequences of these spectra for the top pair production cross section, calculated with the code QQbarThreshold \cite{Beneke:2016kkb} at NNNLO QCD. In all cases, the cross section is affected by initial state radiation, which results in an overall reduction due to events falling below the required center-of-mass energy and a slight broadening of the would-be resonance peak. For the two linear colliders, the beamstrahlung-tail results in a further overall reduction of the cross-section, and in a slight tilt of the plateau region at higher energies. The energy spread of the main luminosity peak leads to a further smearing of the cross-section peak. For FCC-ee, a gaussian luminosity spectrum is assumed, neglecting effects of beamstrahlung, which are substantially smaller than those at linear colliders. This results in an overall larger cross-section and a flatter plateau region. The larger energy spread compared to ILC spreads out the cross section peak slightly more, while the peak is somewhat more pronounced than for the case of CLIC.

\section{The impact of QCD scale uncertainties on the $t\bar{t}$ cross section near threshold}

With the availability of NNLO + NNLL \cite{Hoang:2013uda} and NNNLO \cite{Beneke:2015kwa} QCD calculations of top quark pair production near threshold, the remaining scale uncertainties are now typically in the range of 5\% -- 3\%, fulfilling one of the prerequisites for a precise interpretation of $t\bar{t}$ threshold scans performed with high integrated luminosities. The present study is based on fixed-order NNNLO QCD calculations with additional corrections \cite{Beneke:2015lwa, Beneke:2013kia, Beneke:2010mp}, making use of the code QQbarThreshold \cite{Beneke:2016kkb} with default settings in the PS mass scheme \cite{Beneke:1998rk}, the "native" scheme of the code. 

\begin{figure}[!ht]
\includegraphics[width=0.49\textwidth]{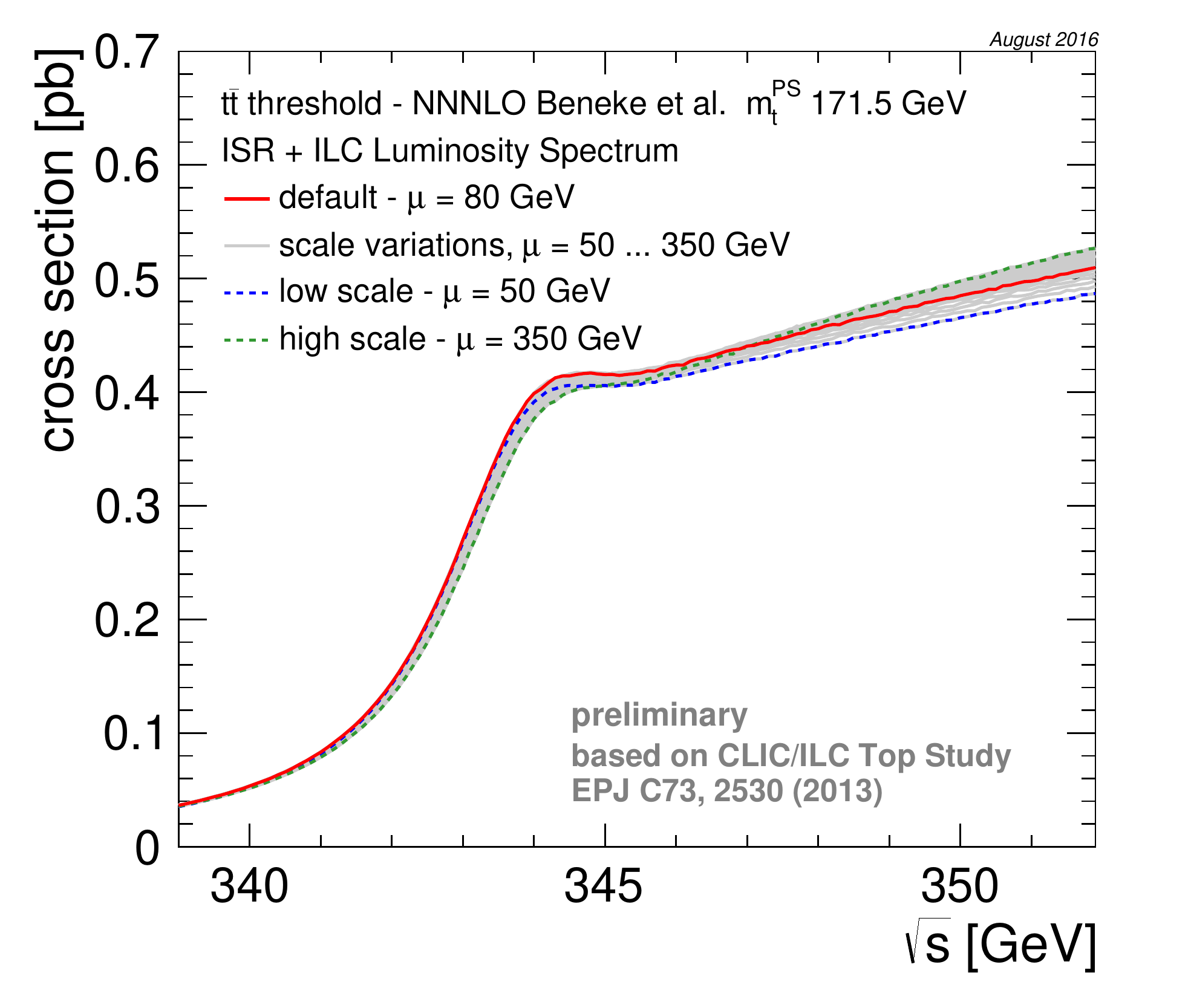}
\hfill
\includegraphics[width=0.49\textwidth]{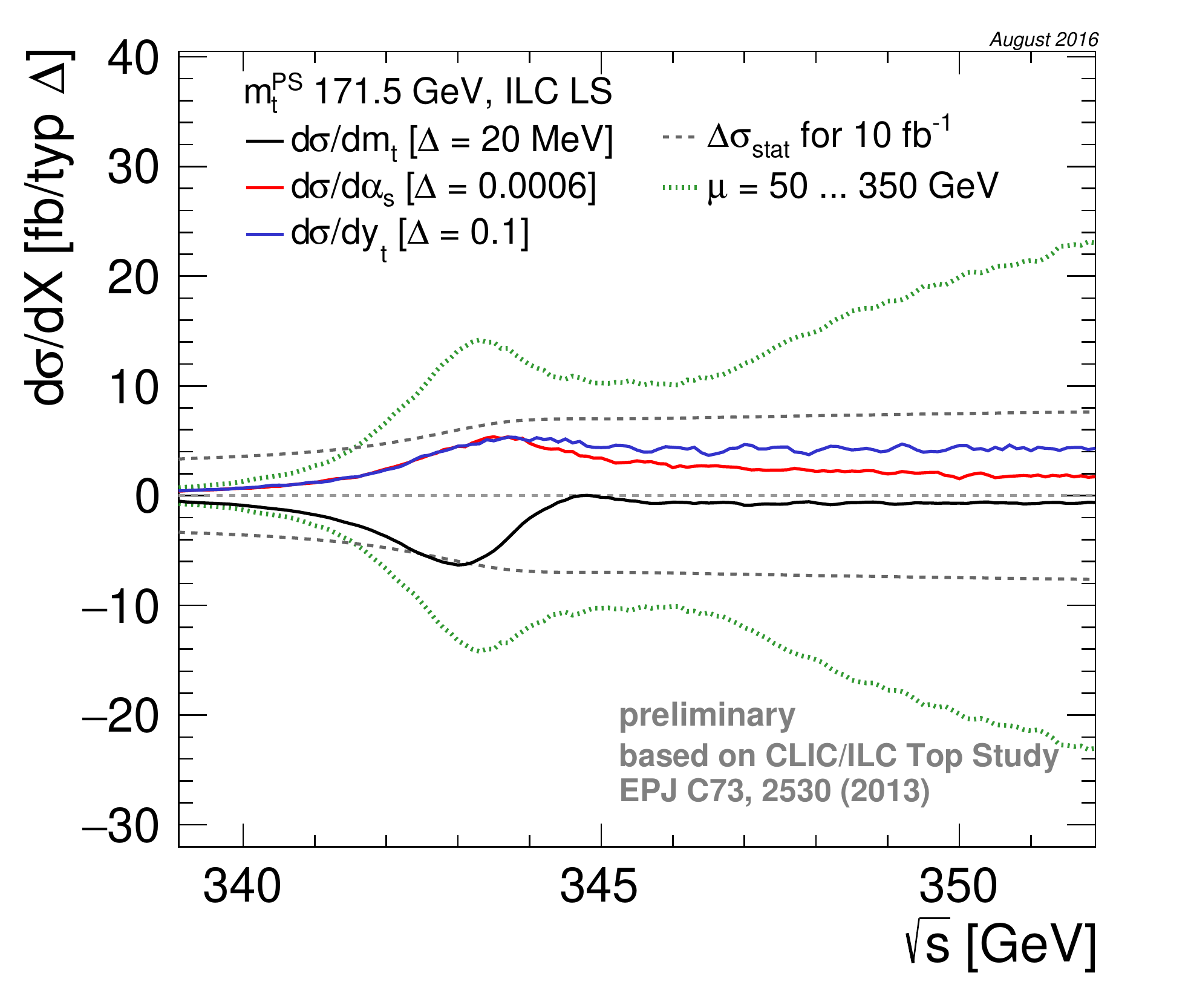}
\caption{The influence of the scale variations on the top cross section ({\it left}) and the size of the cross section uncertainty due to scale variations compared to variations of the cross section for typical expected statistical uncertainties of measurements at a future collider or external uncertainties for a selection of parameters ({\it right}). Also shown are the statistical uncertainties of a 10 fb$^{-1}$ data point. Both figures are shown for the case of ILC. \label{fig:Scale}}
\end{figure}

Figure \ref{fig:Scale} {\it left} illustrates the variation of the top pair production threshold at ILC due to scale variations in the NNNLO calculations. The scale $\mu$ is varied in the range from 50 GeV to \mbox{350 GeV}, assuming a default scale value of 80 GeV. The size of the cross section uncertainty, relative to changes in the cross section induced by changes in the numerical values for selected top parameters is shown in Figure  \ref{fig:Scale} {\it right}. Here, parameter variations corresponding to typical expected statistical uncertainties for future measurements are shown for the mass (20 MeV) and the top Yukawa coupling (0.1), while for the strong coupling 0.0006, the uncertainty of the world average prior to the re-evaluation in Fall 2015 and half of the present uncertainty of the world average \cite{Olive:2016xmw}, is assumed. 

\section{The top mass at threshold accounting for scale uncertainties}

Following the procedures developed in \cite{Seidel:2013sqa} to combine signal efficiencies and background levels obtained from detailed detector simulations with state-of-the-art higher order calculations and the first study of the impact of theory uncertainties on the mass determination via a template fit to the cross section measured in a threshold scan \cite{Simon:2016htt}, the effect of the scale uncertainties on the precision of the top quark mass at ILC, CLIC and FCC-ee are investigated. In the template fit, the $\chi^2$ of the simulated data points to a given top mass template is calculated by considering the distance of the data point to the nearest edge of the template band given by the scale variations. Data points that lie within the band for a given mass thus do not contribute to the overall $\chi^2$ for that particular mass value. 

\begin{figure}[!ht]
\includegraphics[width=0.49\textwidth]{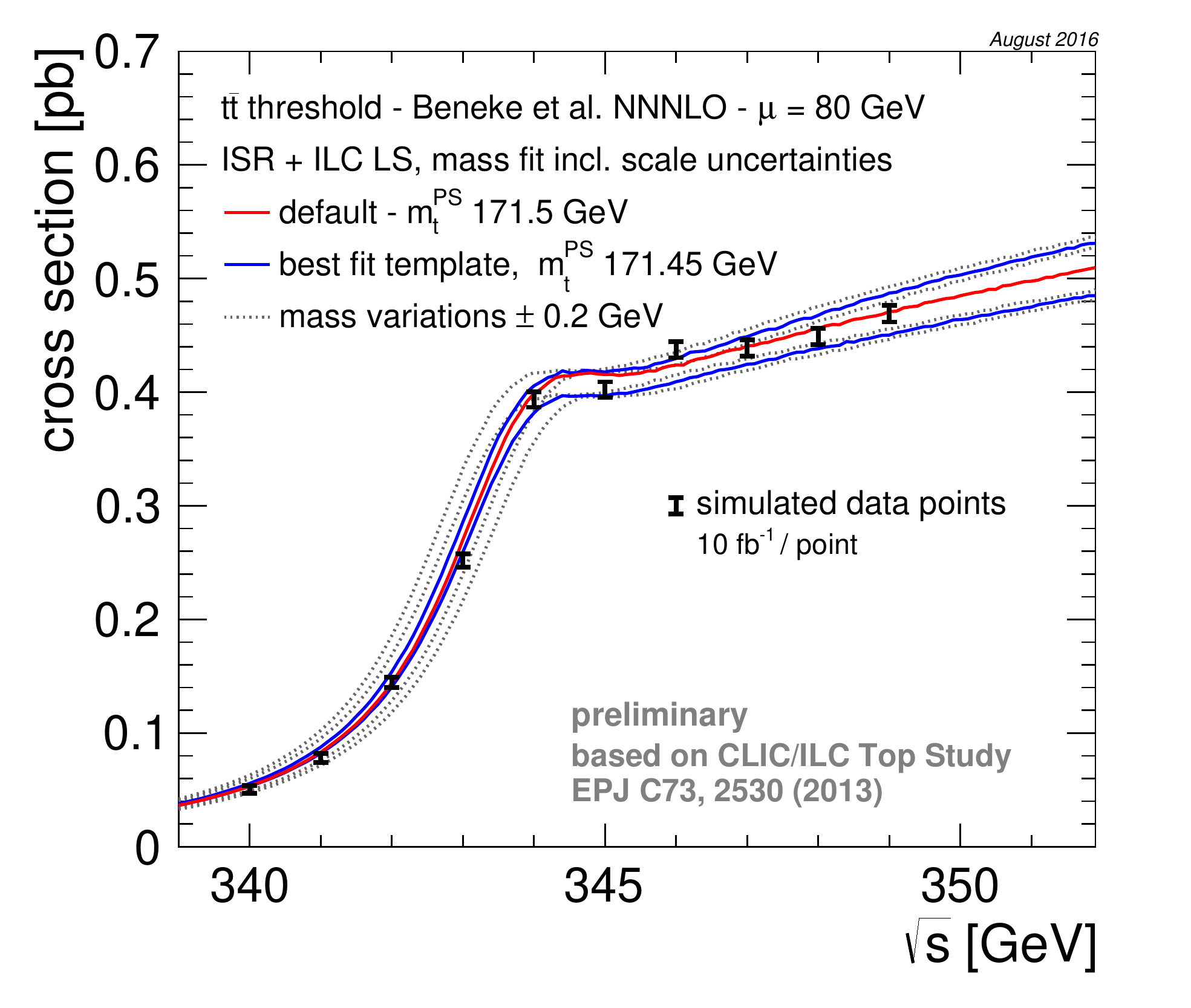}
\hfill
\includegraphics[width=0.49\textwidth]{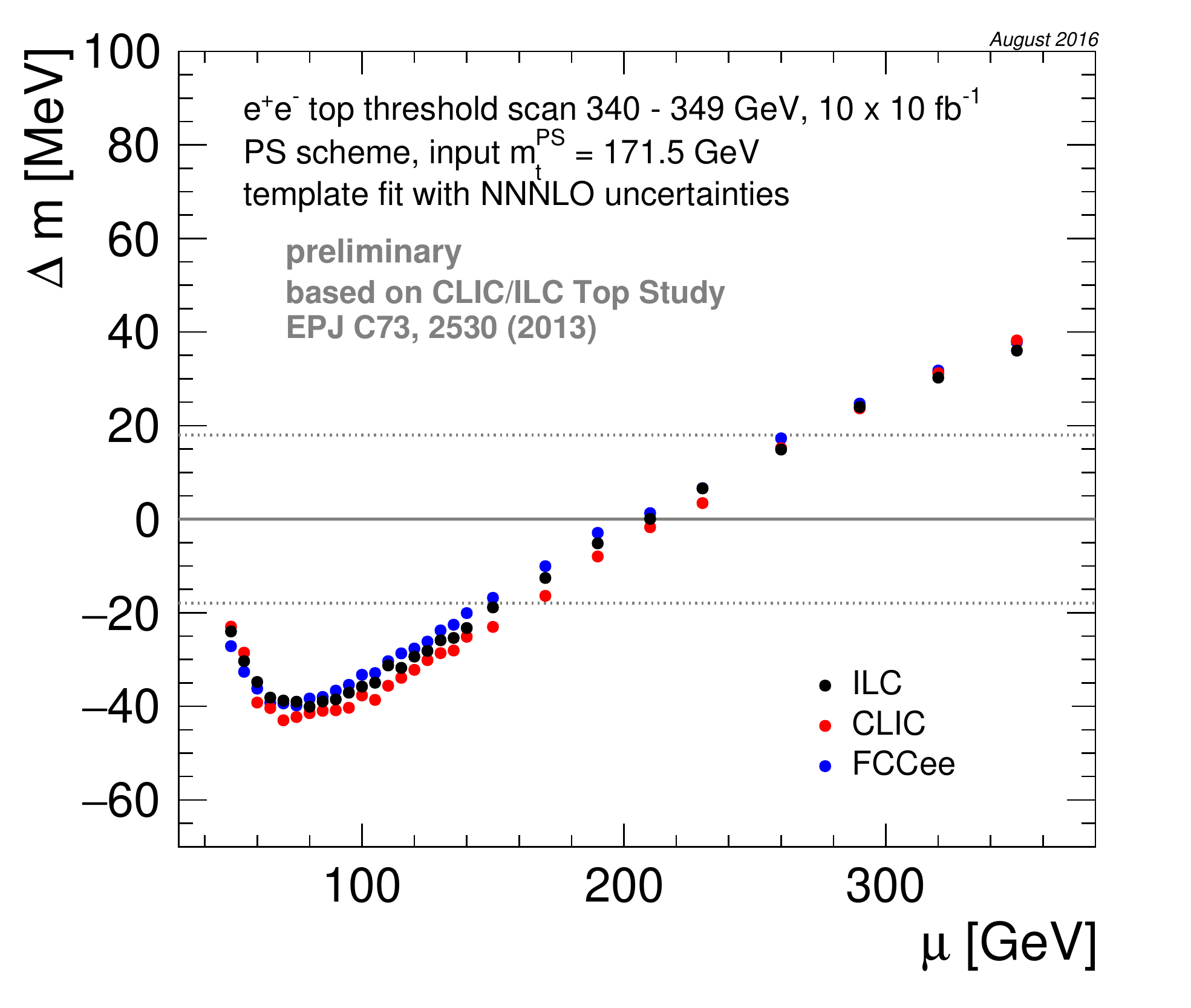}
\caption{Illustration of a threshold scan at ILC with a template fit accounting for scale uncertainties  ({\it left}) and the dependence of the extracted mass depending on the input scale for the three considered collider options, with the statistical uncertainties of ILC illustrated by the dashed horizontal lines ({\it right}). \label{fig:Mass}}
\end{figure}

Figure \ref{fig:Mass} {\it left} shows a simulated threshold scan for the case of ILC, together with fit templates accounting for the cross section uncertainty due to scale variations. The best fit template shows a (trivial) offset of 50 MeV, introduced by the choice of the standard scale which results in the highest cross-section within the range of considered scale values. For an integrated luminosity of \mbox{100 fb$^{-1}$} \mbox{(10 $\times$ 10 fb$^{-1}$)}, the purely statistical uncertainties of the top quark mass in the PS scheme are \mbox{18 MeV}, \mbox{21 MeV} and 15.5 MeV for ILC, CLIC and FCC-ee, respectively. The differences in precision originate from the differences in absolute cross section due to the different level of beamstrahlung-effects as well as from differences in the shape of the threshold curve. As discussed in more detail in \cite{Simon:2016htt}, the use of cross-section bands accounting for the scale uncertainties in the template fit results in larger fit uncertainties for single threshold scans, with a mean fit uncertainty of 28 MeV, 32 MeV and 27 MeV for ILC, CLIC and FCC-ee, respectively. 

Figure \ref{fig:Mass} {\it right} shows the offset of the fitted mass from the input value as a function of the input scale for the simulations for all three collider options. This variation is taken as the mass uncertainty due to the remaining scale uncertainty of the theory cross section. For all three colliders, the variations are essentially identical, and with a range of $\pm$40 MeV they dominate over the statistical uncertainties. 

\section{Conclusions \& Outlook}

Following a first look at the impact of scale uncertainties of NNNLO QCD calculations of top quark pair production on the measurement of the top quark mass in a threshold scan at the International Linear Collider, this contribution extends this preliminary study to an investigation of potential differences for different $e^+e^-$ colliders. While the differences in the luminosity spectrum result in different statistical uncertainties, with an approximately 20\% smaller uncertainty at FCC-ee compared to linear colliders for the same integrated luminosity of 100 fb$^{-1}$, the theory uncertainties due to scale variations considered in the calculation of the cross section are essentially independent of the collider details. In the PS mass scheme, this uncertainty amounts to approximately 40 MeV, substantially larger than the statistical uncertainties. This suggests that theoretical uncertainties will have a strong influence on the overall precision of the top quark mass attainable at a future $e^+e^-$ collider. 

In the future, the present study will be further extended by the addition of the uncertainties of non-resonant electroweak effects once they become available in the theory calculations used here, and will be performed in different mass schemes. For the circular collider case a higher degree of realism could be added by using realistic detector simulations and a more sophisticated description of the FCC-ee luminosity spectrum. However, such changes are not expected to change the overall conclusions in a substantial way.

\end{document}